\documentclass[conference]{IEEEtran}
\usepackage[T1]{fontenc}
\usepackage{amsmath,bm,amsfonts,amssymb,
graphicx,color,mathtools,setspace,
comment,multirow,xfrac,subcaption,caption,balance}
\usepackage{booktabs}
\usepackage{epstopdf}
\usepackage{mathtools}
\usepackage{bigints}
\usepackage{cite}
\usepackage{citesort}
\newcommand*\varhrulefill[1][0.4pt]{\leavevmode\leaders\hrule 
height#1\hfill\kern0pt}

\begin{document}
\title{On the Impact of Spillover Losses in 28 GHz
Rotman Lens Arrays for 5G Applications\vspace{-6pt}}
\author{\IEEEauthorblockN{Muhammad Ali Babar Abbasi, Harsh Tataria, Vincent F. Fusco, and Michail Matthaiou}
\IEEEauthorblockA{Institute of Electronics, Communications and Information Technology (ECIT), 
Queen's University Belfast, Belfast, U.K.}
\IEEEauthorblockA{e--mail:\{m.abbasi, h.tataria, v.fusco, m.matthaiou\}@qub.ac.uk}
\vspace{-23pt}}

\maketitle

\begin{abstract}
This work demonstrates the sensitivity of lens antenna arrays operating at millimeter-wave (mmWave) frequencies. Considering a Rotman lens array in \emph{receive} mode, our investigation focuses on its most imperative defect: \emph{aberration} of electromagnetic (EM) energy. Aberration leads to \emph{spillover} of electric fields to neighboring ports, reducing the lens' ability to focus the EM energy to a desired port. With full EM simulations, we design a 28 GHz, 13 beam and 13 array port Rotman lens array to characterize its performance with the aforementioned impairment. Our findings show that the impact of aberration is more pronounced when the beam angles are close to the \emph{array end-fire}. More critically, the corresponding impact of aberration on the desired signal and interference powers is also investigated for an uplink multiuser cellular system operating at 
28 GHz. The presented results can be used as a reference to re-calibrate our expectations for Rotman lens arrays at mmWave frequencies.
\end{abstract}
\IEEEpeerreviewmaketitle

\vspace{-5pt}
\section{Introduction}
\label{Introduction}
\vspace{-2pt}
Radio-frequency (RF) lens-enabled antenna arrays 
have considerable potential to reduce the hardware 
complexity at millimeter-wave (mmWave) frequencies 
\cite{SAYEED1}.\setlength{\skip\footins}{4pt}
\let\thefootnote\relax\footnote{This 
work was supported by the EPSRC, UK, under grant 
EP/P000673/1 and EP/EN02039/1.} 
Nevertheless, lens arrays suffer 
from the inherent quantization of the beamspace. 
In addition, errors due to imperfections in the 
lens construction itself have a critical impact 
on its performance. Almost all of the existing 
studies on lens-based architectures as a radio 
transmitter or receiver consider the 
lens as an \emph{ideal (perfect) RF device} 
(see e.g., \cite{SAYEED1,TATARIA1,GAO1,ZENG1} 
and references therein). A common topology in the 
literature is known as the Rotman lens, which is 
designed with \emph{three} focal points where lens 
operation is perfect. At all 
other focal points, lens aberration occurs 
\cite{jiang2017metamaterial}, causing 
\emph{spillover losses} - as electromagnetic (EM) 
energy which may be desired for a particular beam 
port also leaks into neighboring beam ports. This 
counteracts the functionality of the lens, causing 
high levels of desired signal loss. The \textit{impact} 
of this physical artifact on the overall system performance remains 
uncharacterized in the literature, and in this paper, we 
close this gap. 
More specifically, with a $13\times{}13$ 
Rotman lens antenna array at 28 GHz, we investigate the 
impact of aberration on an uplink multiuser cellular 
system operating with analog and baseband processing. In doing so, 
via EM simulations, we characterize the inherent 
limitations of Rotman lens arrays by studying the 
spillover levels as a function of the azimuth 
direction-of-arrivals (DoAs). 

\vspace{-4pt}
\section{Design Principles of Rotman Lens Arrays}
\vspace{-1pt}
\label{RotmanDesign}
\begin{figure}[!t]
	\vspace{-5pt}
	\centering
	\includegraphics[width=8cm]{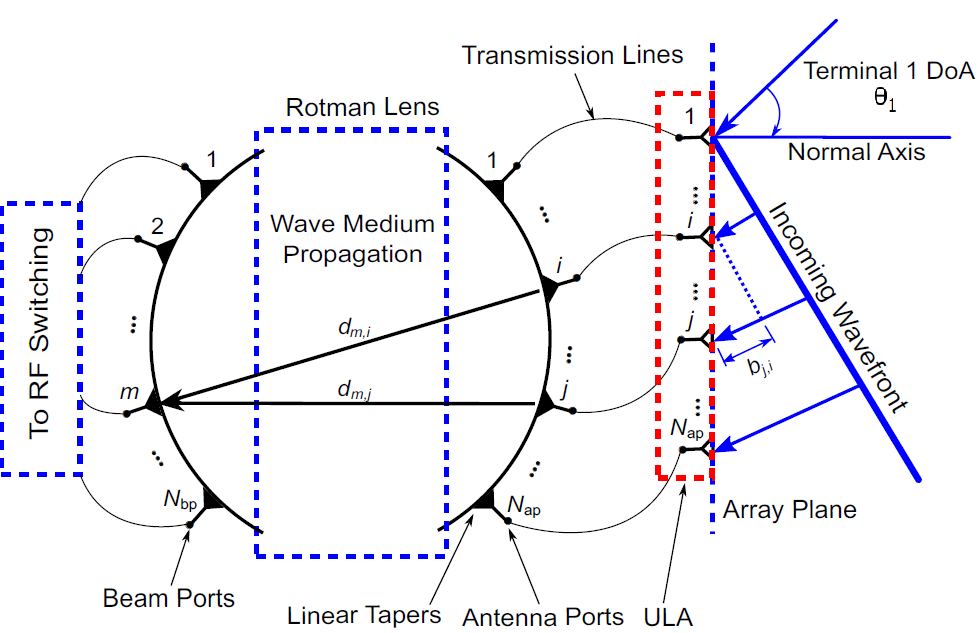}
	\caption{Rotman lens schematic diagram.}
	\label{RotmanLens}
	\vspace{-17pt}
\end{figure}

Successful operation of mmWave cellular systems will 
requires \emph{pencil-like} beam-steering capability in order to accurately 
steer/receive the transmit/receive gain towards/from the strongest 
scatterer(s) in the far-field propagation channel. 
Most commonly, this is done via a 
network of analog phase shifters interfaced with the array elements 
(see e.g., \cite{ROH1}). However, it is well known that 
mmWave phase shifters tend to be 
significantly lossy, and often incapable of providing \emph{precise} phase shifts over the 
required system bandwidth. 
More accurate phase shifters based on tunable materials 
often need to be driven by an external control signal, leading to
higher circuit complexity and material cost \cite{RIAL1}. 
\textit{Rotman lenses} provide an attractive alternative to phase shifter networks, 
which are increasingly considered \cite{SAYEED1,GAO1,TATARIA1,ZENG1}. 
A Rotman lens provides \emph{switch-less} multi-beam 
operation of an antenna array with a wide beam-steering range. 
Figure~\ref{RotmanLens} 
presents the generalized schematic diagram of a Rotman lens with 
$N_{\textrm{bp}}$ beam-ports and $N_{\textrm{ap}}$ array ports, 
all connected to the corresponding tapering and transmission lines. 
The area annotated as \emph{wave medium propagation} is where 
wave superposition takes place. The tapering
and transmissions lines are normally used to \emph{artificially} introduce time 
delays such that the waves propagating through the transmission lines are 
phase aligned along the port terminations. A separate set of ports generally 
referred to as the \emph{dummy ports}, are introduced to increase the adjacent beam 
port isolation. The dummy ports are normally terminated in a matched load, 
and thus result in minimizing the reflections from the side walls of the lens. The beam 
side and array side curvature is defined using design parameters: on-axis 
focal length $f_{1}$, off-axis focal length $f_{2}$, their ratio 
$\beta=f_{2}/f_{1}$, the focal angle $\alpha$, the sweep angle 
$\varphi_{\textrm{max}}$, and the lens expansion factor defined as 
$\gamma=\sin\left(\varphi_{\textrm{max}}\right)/\sin\left(\alpha\right)$. 
A detailed description of the above parameters can be 
found in \cite{HANSEN1}. While realizing a specific Rotman lens geometry, the aforementioned parameters define a convex polygon, where the beam and the array port focal points can be identified \cite{HANSEN1}. The part of the lens around these points is usually referred to as the \emph{port segment}, and is connected to tapered lines which \emph{guide} the propagating wave towards corresponding transmission line. 
\begin{figure}[!t]
	\centering
	\hspace{-15pt}
	\includegraphics[width=5.5cm]{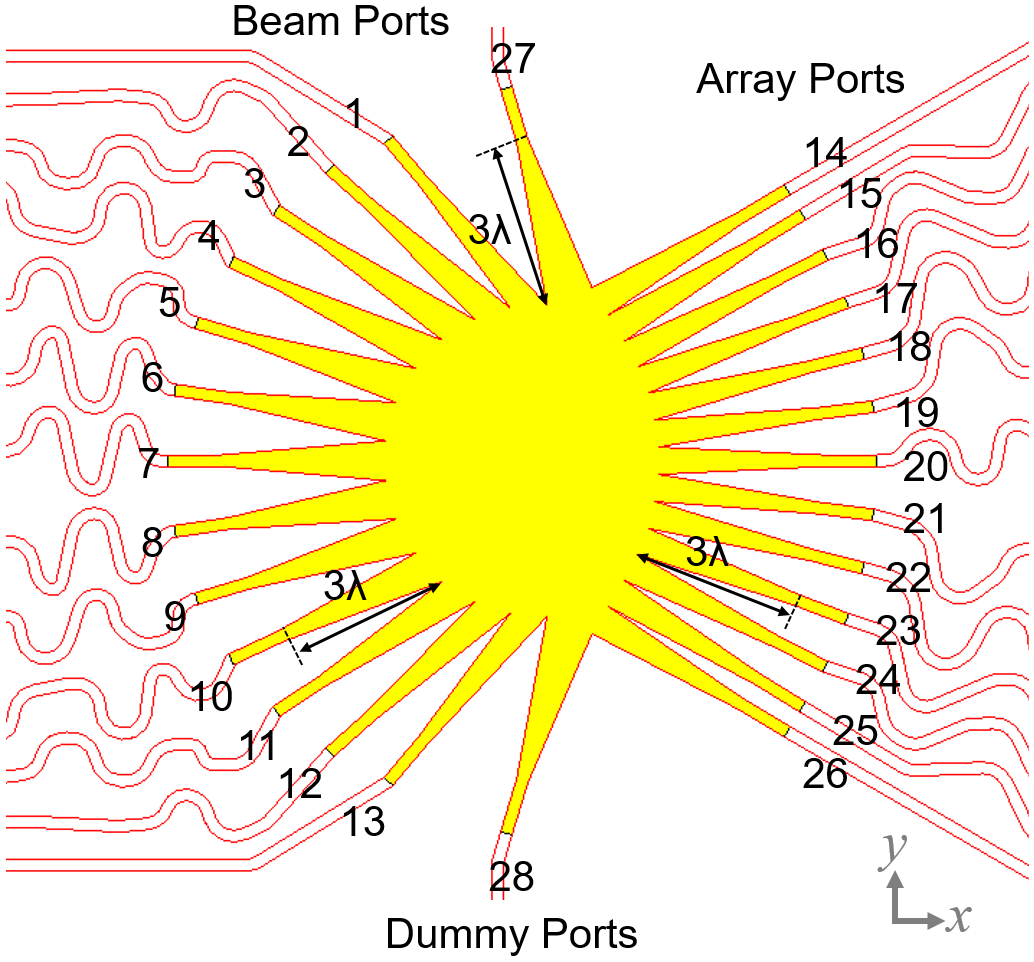}
	\caption{EM simulation of a 28 GHz Rotman lens 
	using FDTD method. The labels 1--13 denote the 
	beam ports, 14--26 denote the array port while 27 and 28 denote 
	the dummy ports.}
	\label{LensPhysical}
	\vspace{-17pt}
\end{figure}

\vspace{-3pt}
\section{28 GHz Electromagnetic Design\\of a Rotman Lens Array}
\label{28GHzDesign}
Here we present the design of a 28 GHz Rotman lens constructed on a 0.64 mm 
thick Taconic-RF 60 substrate ($\epsilon_{\textrm{r}}=6.15, \tan\left(\delta\right)=0.0038)$ using \emph{microstrip} technology. In our design, 
$N_{\textrm{ap}}=N_{\textrm{bp}}=13$, and two dummy ports are 
considered. The lens's parallel plate region was synthesized by the design 
parameters of a tri-focal Rotman lens model \cite{GAO1,HANSEN1}. 
The predefined design parameters were $f_{1}=5\lambda$, 
$\beta=0.9$, $\alpha=30^{\circ}$, 
$\varphi_{\textrm{max}}=30^{\circ}$, and the array steering 
angle $\theta$ was set to $50^{\circ}$. The tapering lines for 
all the ports are $3\lambda$ in length. Moreover, the 
Finite-Difference-Time-Domain (FDTD) method was used 
to characterize the lens in full-wave electromagnetic (EM) 
simulation. The physical lens geometry is presented in 
Fig.~\ref{LensPhysical} with port numbering defined in the 
caption. Although this specific example was constructed using 
standard synthesis method, it is important to mention that with 
a careful selection of the design parameter of the parallel plate 
region, the lens performance can be advanced to an extent. 
Further to this, it is worth mentioning that the geometrical 
parameters of the overall Rotman geometry (parallel plate 
region, dummy ports, tapered transitions and transmission
 lines) can be optimized to further improve the lens 
performance \cite{GAO1}.

\vspace{-3pt}
\section{Results and Discussion}
\label{ResultsandDiscussion}
\begin{figure}[!t]
 \centering
 \begin{subfigure}[b]{0.2\textwidth}
 \centering
 \includegraphics[width=\textwidth]{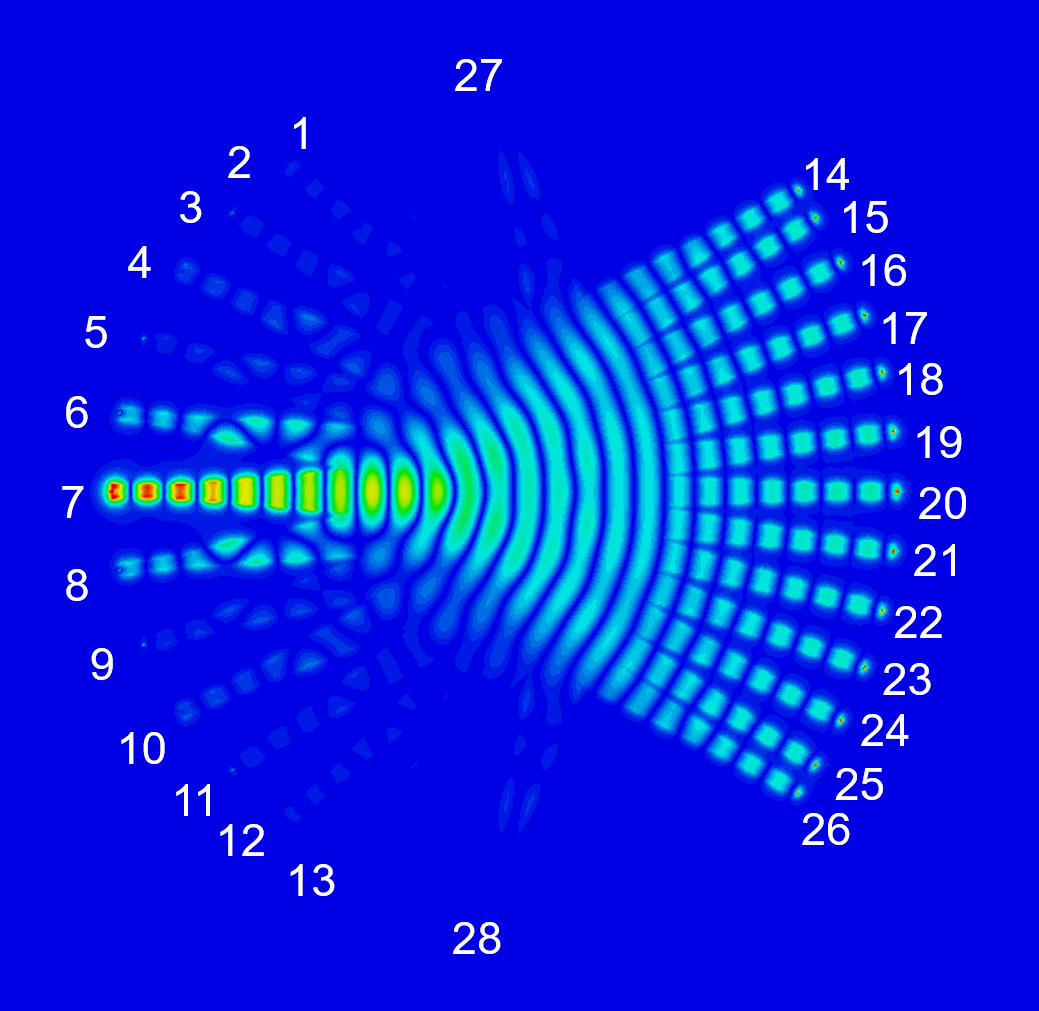}
 \caption{}
 \end{subfigure}\hfill
 \begin{subfigure}[b]{0.2\textwidth}
 \centering		 
		 \hspace{10pt}
 \includegraphics[width=\textwidth]{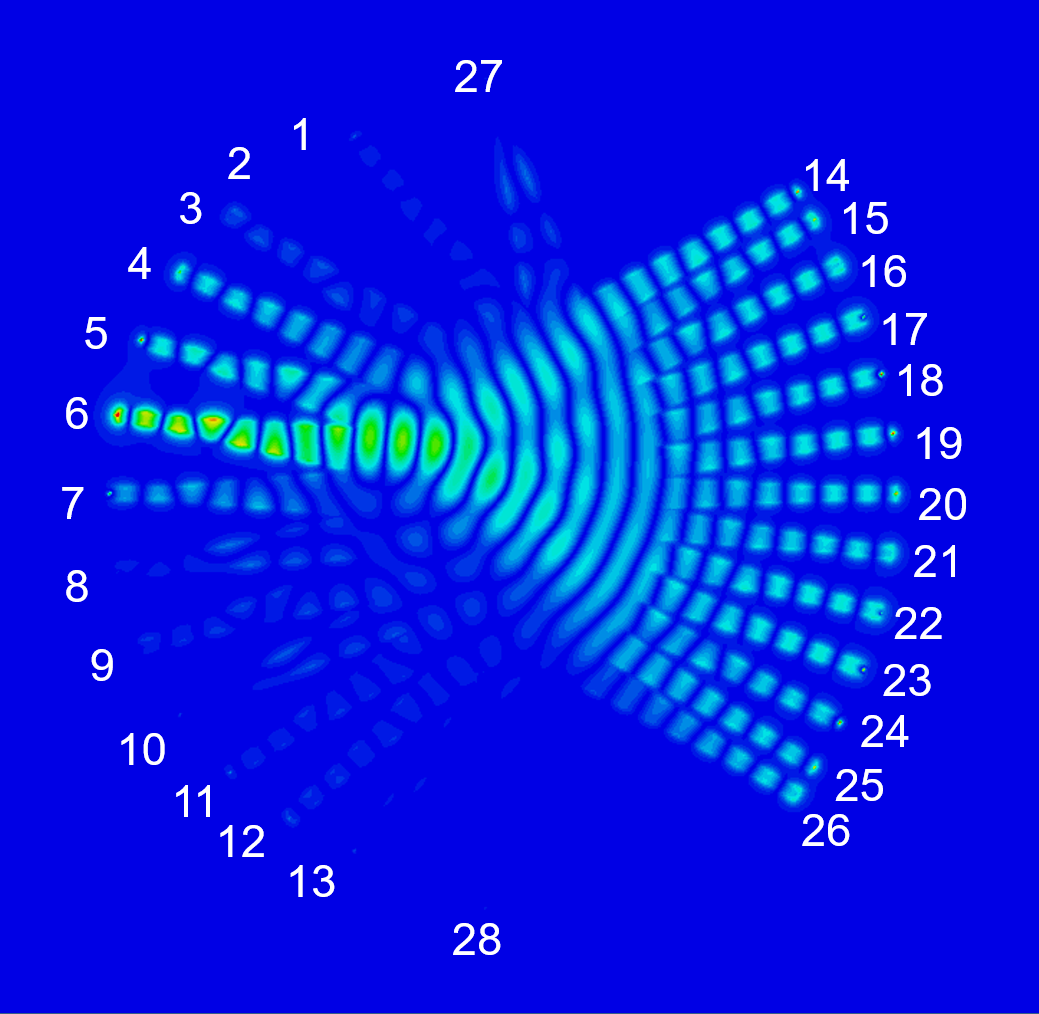}
 \caption{}
 \end{subfigure}\hfill
 \begin{subfigure}[b]{0.2\textwidth}
 \centering
 \includegraphics[width=\textwidth]{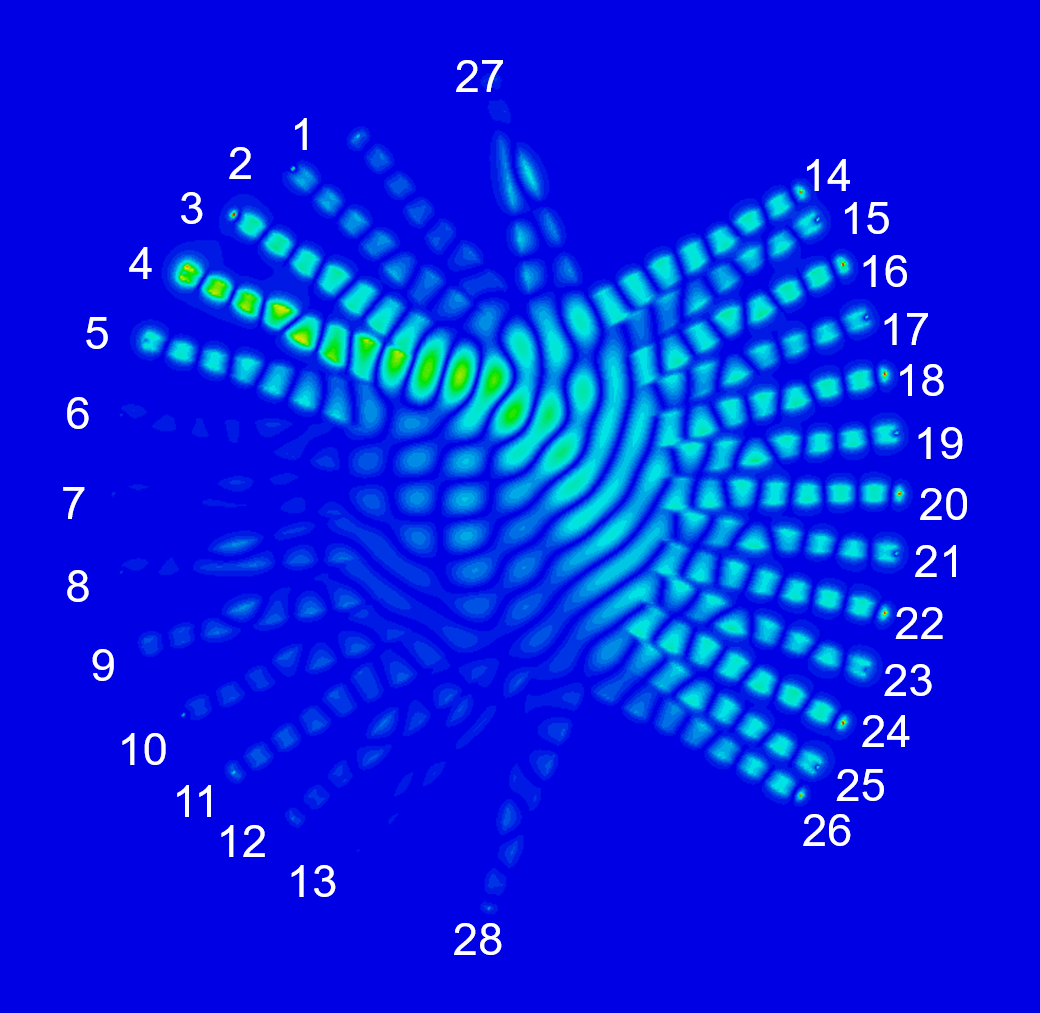}
 \caption{}
 \end{subfigure}
 \vspace{-3pt}
 \caption{Surface E-field distribution $200\mu{}$m 
inside the substrate layer at multiple angle of arrivals at 
(a) $\theta=0^{\circ}$, (b) $\theta=12.5^{\circ}$ and 
(c) $\theta=26.5^{\circ}$ (color map: normalized for all 
figures).}
\label{DOAResults}
\vspace{-19pt}
\end{figure}

\begin{figure}[!t]
	\centering
	\begin{subfigure}[b]{0.17\textwidth}
		\centering
		\includegraphics[width=\textwidth]{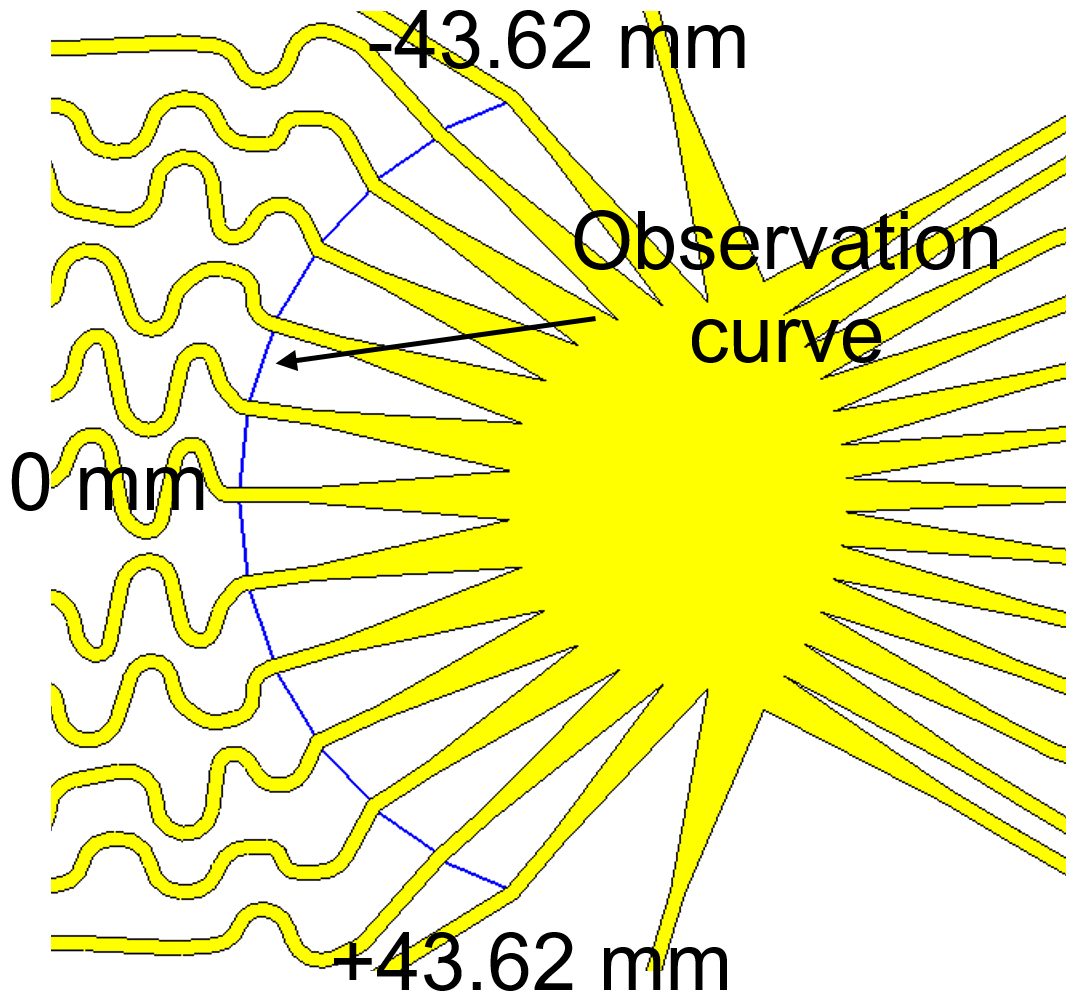}
		\vspace{-13pt}
		\caption{}
	\end{subfigure}\hfill
	\begin{subfigure}[b]{0.4\textwidth}
		\centering	    	 
		\hspace{10pt}
		\includegraphics[width=\textwidth]{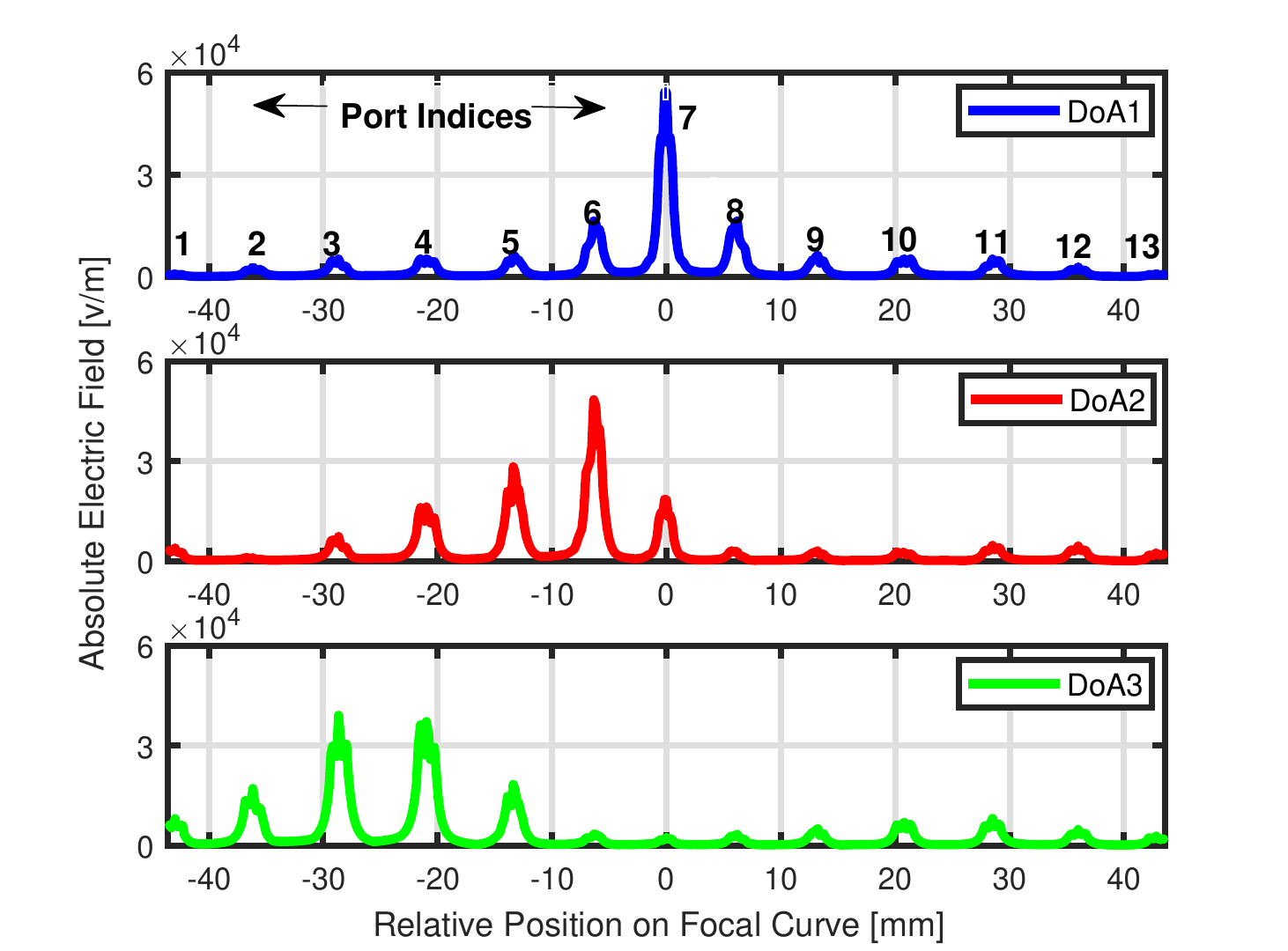}
		\caption{}	
		\vspace{-3pt}
	\end{subfigure}\hfill
	\caption{(a) Indication of the observation curve in the Rotman 
		lens substrate. (b) Electric field distribution along the observation 
		curve depicting the field spillover at multiple DoAs when 
		$\theta=0^{\circ}$, $\theta=12.5^{\circ}$, and 
		$\theta=26.5^{\circ}$.}
	\label{fig4}
	\vspace{-21pt}
\end{figure}

\vspace{-1pt}
A set of EM simulations were carried out in which 
array ports of the lens were excited by phase ramped 
power signals representing multiple DoAs 
along the azimuth plane. For the sake of brevity, only 
\emph{three} distinct scenarios are presented in 
Fig.~\ref{DOAResults}. In the first case 
(denoted by DoA1), the beam is propagating 
in a line-of-sight (LOS) manner from a potential 
transmitting source which could be located 
at $\theta=0^{\circ}$ relative to the 
antenna array (see Fig.~\ref{RotmanLens}). Based on the electric field distributions 
shown in Fig.~\ref{DOAResults} (a), it is evident 
that the \emph{maximum} power is converged at the 
central beam port, i.e., port 7, while a small portion 
of the power is spilled over to the neighboring ports 
due to aberration. One can also notice 
from Figs.~\ref{DOAResults} (b) and (c), that the 
amount of spillover \emph{increases} as $\theta$ is varied 
from $12.5^{\circ}$ (DoA 2) to 
$26.5^{\circ}$ (DoA 3). One can also observe the \emph{reflection} in addition to  
spillover towards the \emph{opposite ports (9 - 12)} in case of DoA 3; 
a trend which is not significant at broadside-like angles. \emph{Note that the 
DoA 2 and DoA 3 cases do not coincide with any 
of the designed focal points of the Rotman lens array, 
and hence are purely demonstrating the spillover 
effects on the Rotman lens focusing.} This result 
highlights one of the major limitations of the Rotman lens, and demonstrates 
the fact that the EM focusing is \emph{more accurate} towards the 
broadside excitation angles. 
To quantify the field leakage into the neighboring ports of 
the lens, we show an observation curve 
$200 \mu{}$m inside the Taconic 
substrate layer (see Fig.~\ref{fig4} (a)). 
The one-dimensional plot of electric field as a function of 
the Rotman curve length for \emph{all} DoA cases 
is presented in Fig.~\ref{fig4} (b). 
The field maxima along the curve length are indicative of 
13 ports. It is interesting to note that the 
field distribution for DoA1 is fairly symmetric. This is 
unlike DoA2, which reveals a contour with an 
\emph{uneven} distribution of fields, one where the spillover 
profile is vastly different for the central and the edge ports. 
The final case in Fig.~\ref{fig4} (b) the contour for 
DoA3, where the wave converging point 
falls at the boundary of two ports. Catastrophically, the 
power is distributed between two concurrent ports almost 
equally. Here the effects of RF reflections can also be 
observed in addition to the spillover, where 
peaks in the contour profile are observed between 
20-40 mm. 

\begin{figure}[!t]
	\vspace{-10pt}
	\centering
	\hspace{-25pt}
	\includegraphics[width=7.6cm]{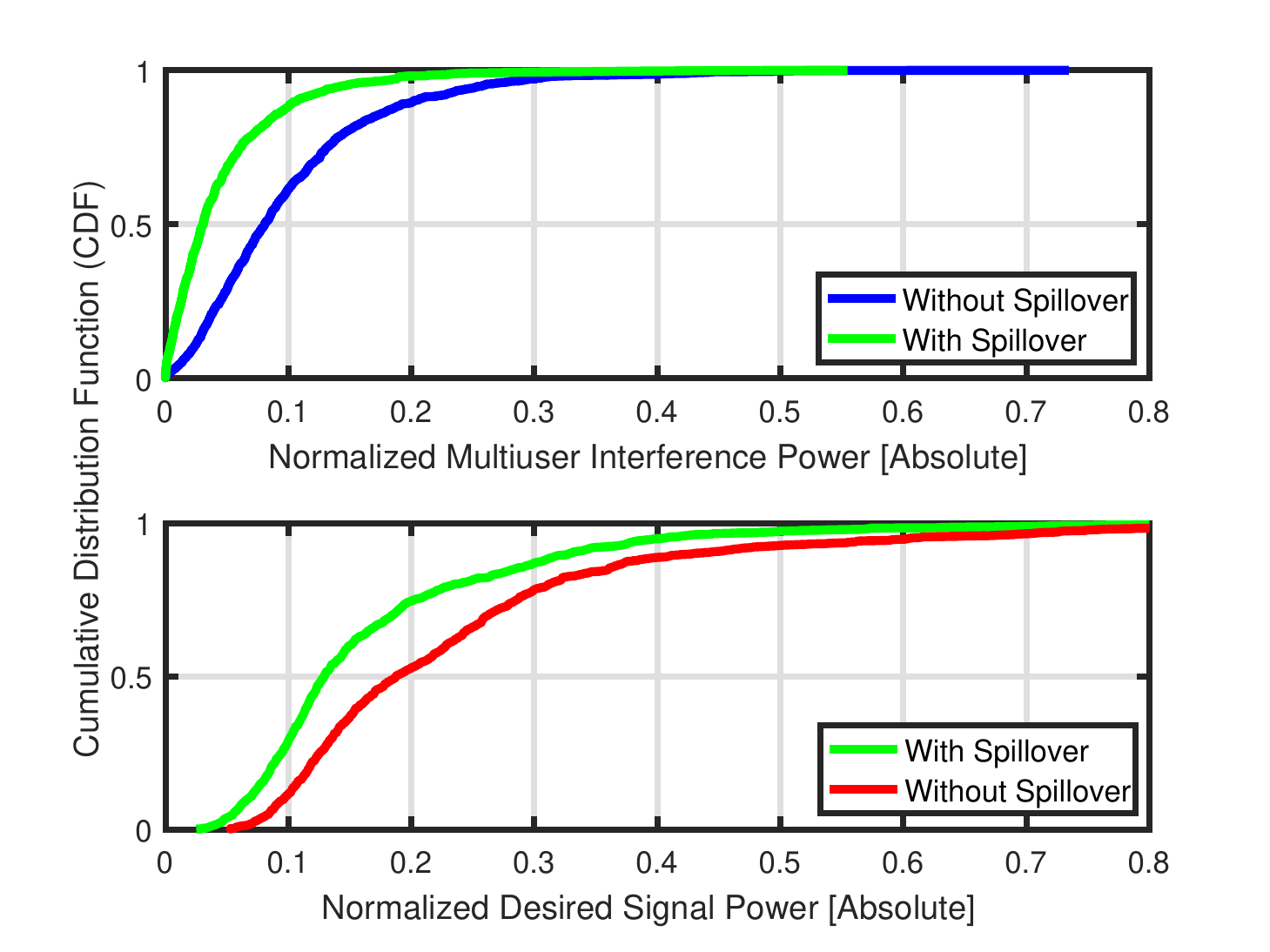}
	\hspace{-18pt}
	\vspace{-3pt}
	\caption{Desired signal and interference powers at a user terminal with 13$\times$13 Rotman lens array receiving signals from two simultaneous users at an average operating SNR of 0 dB. }
	\vspace{-15pt}
	\label{capacity}
\end{figure}

In order for one to understand the ultimate 
impact of aberration, we simulate an uplink multiuser MIMO system, 
where \textit{two} mobile terminals simultaneously transmit uplink data to a 
Rotman lens-enabled array. Upon receiving the signals, the Rotman array is fed with a 
network of RF switches, followed by two complete down-conversion chains to 
recover the transmitted data at baseband. 
We employ the use of maximum-ratio (MR) baseband combining to separate the 
multiple arriving streams. The 28 GHz far-field propagation channel 
is simulated following the classical \emph{double-directional description} 
\cite{MOL}. Precisely, we consider the total number of scattering clusters in the 
propagation channel to be 4, with each cluster contributing the total of 
5 sub-paths. The instantaneous path gains are assumed to be \emph{Gaussian} 
distributed with zero-mean and unit variance. The far-field array steering  
vector of a linear array is utilized with a \textit{uniform} beam scanning 
range in the forward half space of the array ($\theta\in\left[-90^\circ, 90^\circ\right]$). 
Large-scale fading (geometric attenuation and shadow fading/blockage) is modeled on 
both links using 
the classical power loss model described in \cite{TATARIA1}. 
Due to space constraints, 
we avoid presenting further information regarding the simulation setup, as well as 
the exact mathematical descriptions of the desired signal and interference powers with 
MR processing. These can be found in \cite{TATARIA1}. 
Figure \ref{capacity} depicts the desired signal and multiuser interference power 
cumulative distribution functions (CDFs) \textit{with and without} aberration, at an 
operating SNR of 0 dB. One can observe that with aberration, there is naturally a loss 
in the desired signal power, more pronounced at probability 
values around 0.5. This is due to the \emph{leakage} of the desired EM energy to the 
neighboring ports, allowing the power to spread out. 
In stark contrast, a \textit{reduction} in the  
interference power is observed with aberration. This is due to the fact that the spreading of the 
EM energy across multiple ports effectively helps to \textit{de-correlated} the uplink 
signals in the RF domain, such that the interference power is not concentrated on a particular port, but 
rather distributed to a set of ports. It is worth noting that overall, it is the \textit{ratio} of the desired 
signal to the interference power which is seen at a user terminal. The 
dominance of one over the other is a function of \emph{all} involved system 
parameters, as well as aberration levels. Hence, in order not to \textit{obfuscate} 
the findings, we avoid discussing the ratio explicitly.

\vspace{-5pt}
\section{Conclusions}
\label{conclusions}
\vspace{-2pt}
We present an investigation into the 
spillover losses of a 28 GHz Rotman lens array. Via 
the aid of full EM simulations, for a 13$\times$13 
Rotman array, we conclude that the impact of 
spillover is significantly more pronounced at 
beam angles closer to the array end-fire. More 
accuracy in the spatial focusing is found when 
the arriving beam at the Rotman array is around 
the broadside direction, corresponding to exciting 
the central ports. The impact of aberration on the desired signal and 
interference characteristics were investigated. 
To the best of the authors' knowledge, 
such a type of investigation is missing 
from the lens literature. 

\vspace{-6pt}
\bibliographystyle{IEEEtran}

\end{document}